\documentclass[11pt]{article}

\usepackage[preprint]{acl}
\usepackage{amsmath} 
\usepackage{color}
\usepackage{stfloats}  
\usepackage{enumitem}

\usepackage{booktabs, multirow, graphicx, xcolor, colortbl, rotating, pifont}
\newcommand{\cmark}{\ding{51}}
\newcommand{\xmark}{\ding{55}}
\usepackage{times}
\usepackage{latexsym}
\usepackage{booktabs, array, longtable, xcolor}
\usepackage[T1]{fontenc}

\usepackage[utf8]{inputenc}

\usepackage{microtype}

\usepackage{inconsolata}

\usepackage{graphicx}

\usepackage{booktabs}
\usepackage{multirow}
\usepackage{multicol}
\usepackage{longtable}

\usepackage[dvipsnames,svgnames]{xcolor}
\usepackage[many]{tcolorbox} 

\definecolor{main}{HTML}{222222}    
\definecolor{sub}{HTML}{dddddd}     

\tcbset{
    sharp corners,
    colback = white,
    before skip = 0.2cm,    
    after skip = 0.5cm      
}                           

\newtcolorbox{boxD}{
    colback = sub, 
    colframe = main, 
    boxrule = 0pt, 
    toprule = 3pt, 
    bottomrule = 3pt 
}

\usepackage{listings}
\usepackage{fontawesome5}
\usepackage{hyperref}

\title{
MM-tau-p$^2$: Persona-Adaptive Prompting for Robust Multi-Modal Agent Evaluation in Dual-Control Settings
}

\author{
  Anupam Purwar\thanks{Corresponding author: \texttt{anupam.aiml@gmail.com}. Project page: \url{https://anupam-purwar.github.io/page/}.}
  \and
  Aditya Choudhary\footnotemark[1] \\
  Sprinklr AI \\
  Gurugram, India \\ \\
  \faGithub \href{https://github.com/anupam-purwar/mm-tau-p2}{~~Code}}

\begin{document}

\maketitle

\begin{abstract}
Current evaluation frameworks and benchmarks for LLM powered agents focus on text chat driven agents, these frameworks do not expose the persona of user to the agent, thus operating in a user agnostic environment. Importantly, in customer experience management domain, the agent's behaviour evolves as the agent learns about user personality. With proliferation of real time TTS and multi-modal language models, LLM based agents are gradually going to become multi-modal. Towards this, we propose the MM-tau-p$^2$ benchmark with metrics for evaluating the robustness of multi-modal agents in dual control setting with and without persona adaption of user, while also taking user inputs in the planning process to resolve a user query. In particular, our work shows that even with  state of-the-art frontier LLMs like GPT-5, GPT 4.1, there are additional considerations measured using metrics viz. multi-modal robustness and turn overhead  while introducing multi-modality into LLM based agents. Overall, MM-tau-p$^2$ builds on our prior work FOCAL and provides a holistic way of evaluating multi-modal agents in an automated way by introducing 12 novel metrics. We also provide estimates of these metrics on the telecom and retail domains by using the LLM-as-judge approach using carefully crafted prompts with well defined rubrics for evaluating each conversation.
\end{abstract}

\section{Introduction}
Multi-modal LLM based agents are becoming widespread as better voice architectures, 
both open source and closed source, become easily available~\cite{zhang-etal-2024-mm, 
chu2024qwen2audio, palaskar-etal-2024-multimodal, ma-etal-2025-caution, 
durante2024agentai}. Furthermore, LLM based agents with understanding of  the user context are becoming the demand in the customer experience management domain. However, the current benchmarks for LLM based agents have been designed to evaluate their efficacy in responding to user queries, come up with a plan to solve the problem and solve the problem by  performing actions/tool calls with limited or no contextual understanding of user. For example, in tau-bench single-control environment was simulated,  the LLM based agent interacted with the world but the (simulated) user was limited to providing information about preferences and goals. Both tau-bench and tau$^2$-bench explored  retail, telecom and airline domains to evaluate the agent’s ability to solve constraint satisfaction tasks but did not model the expert-novice gap inherent to customer support tasks \cite{yao2024tau, barres2025tau2}. For example, an LLM based agent, while interacting with a user, needs to understand the user’s understanding of the domain, persona and accordingly adapt its response in a dual control setting. Besides, existing benchmarks do not capture the efficacy of agent under multi-modal input viz. text and voice. This gap is evident in other agent benchmarks such as AgentBench  \citet{liu2023agentbench} and WebArena \citet{zhou2023webarena}, which treat users as static oracles rather than dynamic participants with varying levels of domain expertise.  Other existing benchmarks such as VoiceAgentBench \citet{jain2025voiceagentbench} focus on different facets of agent competence viz. speech-driven tool orchestration across multilingual spoken inputs and adversarial cases by VoiceAgentBench. However, none of these benchmarks systematically evaluate multi-modal agents that integrate speech and other modalities with persona adaptation and user-influenced planning. Recent work on FullDuplex conversational agents such as PersonaPlex \citet{roy2026personaplex} evaluated on FullDuplexBench \citet{roy2025fullduplexbench} has made significant progress in natural voice interaction: modeling turn-taking, interruptions, and back-channeling. However, these benchmarks primarily evaluate conversational dynamics and task adherence under a uniform user model. Crucially, they do not evaluate the agent's ability to infer and adapt to the user's level of domain expertise, a capability central to real-world customer support. In contrast to prior approaches, refer Table~\ref{tab:mmtau_summary}, our work directly addresses these gaps by introducing MM-tau-p$^2$, a new benchmark suite that measures multi-modal robustness, persona-adaptive behavior, and planning efficiency in dual-control settings. The notation $p^2$ emphasizes two prompt-centric levers that are necessary for stable evaluation and strong agent performance in dual-control multi-modal settings. The first $p$ denotes \textbf{better prompts for the LLM agent and the human simulator}, including guardrails that prevent premature escalation and encourage the agent to attempt additional reasonable steps before giving up, while keeping the simulator behavior consistent and non-adversarial beyond the intended persona. The second $p$ denotes \textbf{the detailed injection of personas for the user or human simulator}, which controls the level of expertise, the ambiguity, the verbosity and the tolerance to clarification so that the adaptation of the persona can be explicitly measured rather than treated as an uncontrolled artifact.
Unlike prior agent evaluations, MM-tau-p$^2$  captures the evolving interaction between agent and user persona. It uses LLM-as-judge metrics to assess performance degradation across telecom and retail domains. Our results demonstrate that even frontier language models exhibit measurable robustness and efficiency trade-offs when transitioning from text-only to multi-modal. This difference is further amplified in persona-adaptive environments such as customer experience management settings, underscoring the need for holistic agent evaluation beyond existing benchmarks.

\definecolor{oursrow}{RGB}{235,240,255}
\definecolor{headerblue}{RGB}{220,230,255}

\begin{table*}[t]
\label{tab:mmtau}
\centering
\resizebox{\textwidth}{!}{%
\begin{tabular}{llllccccccc}
\toprule
\multirow{2}{*}{\textbf{Benchmark}} &
\multirow{2}{*}{\textbf{Reference}} &
\multirow{2}{*}{\textbf{Domain}} &
\multirow{2}{*}{\textbf{Modality}} &
\multicolumn{7}{c}{\textbf{Capabilities}} \\
\cmidrule(lr){5-11}
& & & &
\rotatebox{60}{\textbf{Dual Control}} &
\rotatebox{60}{\textbf{Tool Use}} &
\rotatebox{60}{\textbf{Multi-Turn}} &
\rotatebox{60}{\textbf{Voice/Speech}} &
\rotatebox{60}{\textbf{Expert-Novice Gap}} &
\rotatebox{60}{\textbf{Persona Adaptation}} &
\rotatebox{60}{\textbf{CX Domain}} \\
\midrule

$\tau$-bench
  & \citet{yao2024tau}
  & Retail, Airline
  & Text
  & \xmark & \cmark & \cmark & \xmark & \xmark & \xmark & \xmark \\

$\tau^2$-bench
  & \citet{barres2025tau2}
  & Retail, Airline, Telecom
  & Text
  & \cmark & \cmark & \cmark & \xmark & \xmark & \xmark & \cmark \\

AgentBench
  & \citet{liu2023agentbench}
  & OS, DB, Web, Game
  & Text
  & \xmark & \cmark & \cmark & \xmark & \xmark & \xmark & \xmark \\

WebArena
  & \citet{zhou2023webarena}
  & E-commerce, Forum, CMS
  & Text
  & \xmark & \cmark & \xmark & \xmark & \xmark & \xmark & \xmark \\

GAIA
  & \citet{mialon2023gaia}
  & General QA
  & Text + MM
  & \xmark & \cmark & \xmark & \xmark & \xmark & \xmark & \xmark \\

VoiceBench
  & \citet{chen2024voicebench}
  & General QA
  & Speech
  & \xmark & \xmark & \xmark & \cmark & \xmark & \xmark & \xmark \\

VoiceAgentBench
  & \citet{jain2025voiceagentbench}
  & General (Multilingual)
  & Speech
  & \xmark & \cmark & \cmark & \cmark & \xmark & \xmark & \xmark \\

FullDuplexBench
  & \citet{roy2025fullduplexbench}
  & General QA
  & Speech
  & \xmark & \xmark & \cmark & \cmark & \xmark & \xmark & \xmark \\

ServiceDuplexBench
  & \citet{roy2026personaplex}
  & Customer Service
  & Speech
  & \xmark & \xmark & \cmark & \cmark & \xmark & \xmark & \cmark \\

\midrule
\rowcolor{oursrow}
\textbf{MM-$\tau$-p (Ours)}
  & ---
  & \textbf{Retail, Telecom}
  & \textbf{Text + Speech}
  & \cmark & \cmark & \cmark & \cmark & \cmark & \cmark & \cmark \\

\bottomrule
\end{tabular}
}
\caption{
  Comparison of related LLM-based agent benchmarks across key dimensions.
  \textbf{Dual Control}: both agent and user can take actions;
  \textbf{Expert-Novice Gap}: benchmark models varying user domain expertise;
  \textbf{Persona Adaptation}: agent adapts responses based on inferred user persona;
  \textbf{CX Domain}: benchmark targets customer experience / support settings.
  MM-$\tau$-p$^2$ is the only benchmark that evaluates all seven capabilities jointly.
  MM = Multimodal; Context Augmentation; DB = Database; CMS = Content Management System; CX = Customer Experience.
}
\label{tab:mmtau_summary}
\end{table*}

\begin{figure*}[htbp]
        \centering
        \includegraphics[width=0.8\textwidth]{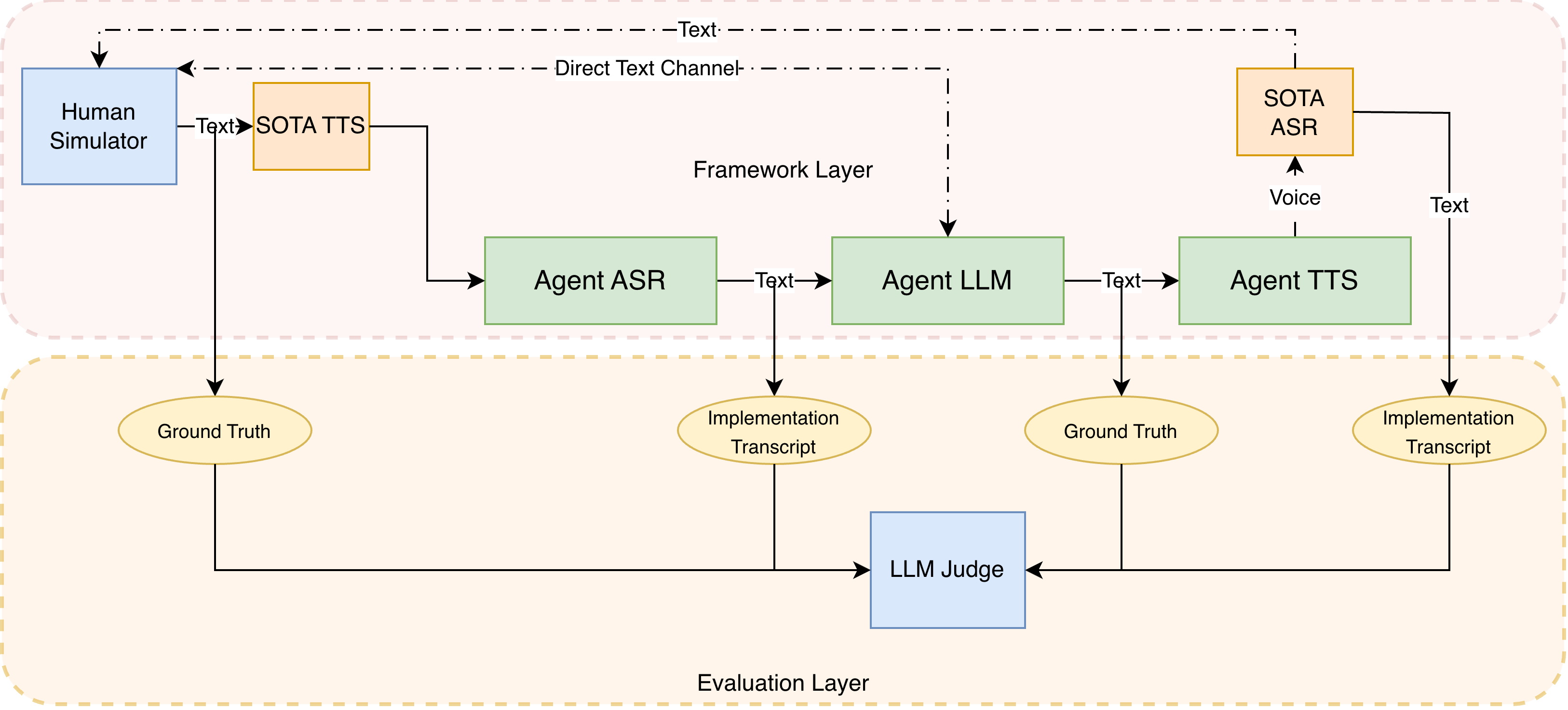}
        \caption{End-to-End Pipeline with conditional edges based on usage (automated vs human-involvement). Automated usage
involves use of SOTA TTS and ASR modules for interaction with the Human-Simulator. Texts are tapped into at various stages
of the pipeline to generate Ground-Truth and Implementation Transcripts (indicated in yellow)}
        \label{fig:pipeline}
\end{figure*}

\section{Methodology}
\label{sec:methodology}

MM-tau-p$^2$ is a benchmark for evaluating multi-modal (voice plus text) LLM agents in customer-support scenarios where the user is an active participant and the agent may or may not have access to user persona information. The benchmark quantifies how agent quality changes when speech interfaces are introduced and when user behavior varies with persona and expertise. MM-tau-p$^2$ supports controlled comparisons across modality and persona conditions, and it reports a suite of automated metrics covering robustness, safety, efficiency, and recovery.

\subsection{Benchmark factors}
\label{sec:factors}

Each evaluation instance in MM-tau-p$^2$ is defined by a tuple
(domain, task, modality, persona, persona, exposure).

\paragraph{Domains and tasks.}
We consider two customer-support domains, \textbf{Telecom} and \textbf{Retail}. Each task specifies a user goal, a set of allowable agent actions (for example, tool calls or API operations), and a task-specific success criterion used to compute pass rates.

\paragraph{Critical fields.}
Each task also defines \textbf{critical fields}, which are error-sensitive entities that can invalidate the outcome if incorrect. Examples include phone number, account identifier, order identifier, amounts, dates, and plan identifiers, depending on the domain. These fields are used by metrics that target catastrophic failure risk.

\paragraph{Modalities.}
We evaluate two modality settings: (i) \textbf{text only}, where the user and agent exchange text messages, and (ii) \textbf{voice only}, where interaction occurs through speech and intermediate transcripts are logged for scoring.

\paragraph{Personas and persona exposure.}
MM-tau-p$^2$ includes three persona settings: \textbf{None} (neutral baseline), \textbf{Easy} (higher domain familiarity), and \textbf{Hard} (lower domain familiarity with more ambiguity and misunderstandings) across different tasks. We evaluated two persona exposure conditions: \textbf{persona injection} (persona metadata provided to the agent) and \textbf{no persona injection} (persona withheld, so the agent must infer user traits from interaction).

\subsection{Dual-control interaction protocol}
\label{sec:dualcontrol}

MM-tau-p$^2$ uses a \textbf{dual-control} protocol where both the user and the agent influence the dialogue. The user may provide incomplete information, correct the agent, request alternatives, or introduce new constraints mid-conversation. This setting tests whether the agent can plan collaboratively, ask clarifying questions appropriately, and maintain efficiency without compromising safety.

\subsection{Voice pipeline and logging}
\label{sec:multimodal}

For inputs based on voice / text evaluation, MM-tau-p$^2$ uses a modular pipeline:
user speech $\rightarrow$ ASR transcript $\rightarrow$ LLM agent $\rightarrow$ agent text $\rightarrow$ TTS speech.
We log (i) the ASR transcript presented to the agent, (ii) the agent outputs and any tool calls, and (iii) the final agent text that is synthesized by TTS. Logging these artifacts supports metric attribution across speech understanding (ASR related), reasoning and planning (LLM related), and response delivery (TTS related).

\subsection{Persona modeling}
\label{sec:persona}
Persona controls how the simulated user communicates, including domain expertise, ambiguity, and tolerance for clarification. In the Easy persona, users more often provide structured information and use correct domain terminology. In the Hard persona, users more often omit key details, use vague language, or provide inconsistent identifiers, reflecting novice behavior in customer support. Comparing persona injection to no persona injection isolates the effect of persona exposure on task success, efficiency, and robustness.

\subsection{User Context Injection}

Apart from the persona, we experiment with providing the agent an inferred personality of the user from the messages in the conversations. Such a context injection enables handling unexpected shifts (such as frustration developed due to agent failures) in user persona over the course of the conversation. For context injection, we pass the conversation history containing the last 16 messages (both user and agent) to the same LLM which is used by the agent and generate personality for the user. The prompt given to the LLM involves basic rubrics for the LLM to classify (refer \ref{context_injection_prompt}) the user on. The rubrics include age, gender, technical skill, articulation skill etc. This provides the agent with degree of to come up with a comprehensive context of the user. This context is injected into the agent's system prompt. The context injection pipeline runs whenever 3 new user messages are available after the last cache update.

\subsection{Automated scoring with LLM-as-judge}
\label{sec:judge}

MM-tau-p$^2$ uses an LLM-as-judge approach to score conversations with rubric-based prompts. The judge assigns turn-level and conversation-level labels using explicit definitions and counting rules. To reduce hindsight bias, labels for a turn are determined only from the conversation context available up to that turn.

\subsection{Evaluation Metrics}

We introduce a comprehensive suite of metrics to evaluate multimodal agent performance:

\paragraph{Goal Achievement Metrics}
\begin{itemize}
    \item \textbf{Critical Field Accuracy (CFA)}: Measures accuracy on error-sensitive fields (e.g., order ID, destination city, passenger name). Computed as $\frac{\text{correct critical fields}}{\text{total critical fields}}$. A single wrong critical field can invalidate task success.
    
    \item \textbf{ARGA (ASR-Robust Goal Achievement)}: Probability of achieving the goal despite ASR errors, $P(\text{Pass}^k \mid \text{ASR error occurred})$. Measures recovery capability rather than ASR accuracy.
    
    \item \textbf{Modality Robustness Score (MRS)}: Degradation from text to voice, computed as $\frac{\text{Pass}^k_{\text{voice}}}{\text{Pass}^k_{\text{text}}}$. MRS $= 1.0$ indicates no degradation; MRS $< 0.7$ suggests the agent is not voice-ready.
\end{itemize}

\paragraph{Efficiency Metrics}
\begin{itemize}
    \item \textbf{Turn Efficiency (TE)}: Ratio of optimal to actual turns, $\frac{T_{\text{optimal}}}{T_{\text{actual}}}$. Values closer to 1.0 indicate efficient task completion without unnecessary back-and-forth.
    
    \item \textbf{Turn Overhead (TO)}: Extra turns incurred in voice vs.\ text interactions, $\frac{T_{\text{voice}} - T_{\text{text}}}{T_{\text{text}}}$. TO $< 0.2$ is minimal; TO $> 0.5$ indicates excessive voice friction.
    
    \item \textbf{User Effort Score (UES)}: Count of user repetitions, corrections, or restatements (e.g., ``No, I said Boston,'' spelling out names). High UES signals poor user experience even when Pass$^k$ succeeds.
\end{itemize}

\paragraph{Recovery Metrics}
\begin{itemize}
    \item \textbf{Error Recovery Rate (ERR)}: Proportion of errors successfully recovered, $\frac{\text{recovered errors}}{\text{total errors}}$. Covers ASR misrecognitions, tool failures, and wrong agent actions recovered via clarification, retry, or undo.
    
    \item \textbf{Recovery Turn Count (RTC)}: Average turns needed to recover from an error, $\frac{\sum \text{recovery turns}}{\sum \text{recovered errors}}$. Domain-specific targets: Retail $\leq 2$, Airline $\leq 3$, Telecom $\leq 2$.
\end{itemize}

\paragraph{Clarification Metrics}
\begin{itemize}
    \item \textbf{Clarification Precision (CP)}: Whether clarifications are necessary, $\frac{\text{necessary clarifications}}{\text{total requested}}$. Low CP indicates over-clarifying when context was sufficient.
    
    \item \textbf{Clarification Recall (CR)}: Whether needed clarifications are requested, $\frac{\text{clarifications requested}}{\text{clarifications required}}$. Low CR means proceeding on ambiguous inputs (e.g., ``Boston'' vs.\ ``Austin'').
\end{itemize}

\paragraph{Safety Metrics}
\begin{itemize}
    \item \textbf{Irreversible Action Safety (IAS)}: Proportion of high-risk actions (cancellations, charges, plan changes) executed only after explicit confirmation, $\frac{\text{confirmed irreversible}}{\text{total irreversible}}$. Must be $\approx 1.0$; any unconfirmed irreversible action is a critical failure.
    
    \item \textbf{Safety Recall (SR)}: Consistency of requesting confirmation when required (e.g., low ASR confidence, ambiguity), $\frac{\text{confirmations requested}}{\text{confirmation-required cases}}$. Ensures no irreversible action bypasses required confirmation.
\end{itemize}

\section{Results}
In our primary experiments, we evaluated frontier LLM agents powered by LLM (GPT 4.1, GPT-5) under MM-tau-p$^2$ in telecom and Retail tasks, comparing text only to voice, and comparing persona injection to non-persona injection. We report metrics aggregated across tasks within each setting and analyze how modality and persona conditions affect robustness, efficiency, and safety.

We evaluate agent performance across Telecom and Retail domains using GPT-4.1 and GPT-5 based LLM-as-judge.
\citet{microsoft2026gpt5vs41} describe GPT-4.1 as optimized for high-speed, high-throughput tasks like real-time 
customer support, while GPT-5 targets complex multi-step reasoning and agentic 
workflows at higher latency~\cite{openai2026models,microsoft2026gpt5vs41}. Using both 
models as judges enable comparison whether evaluation outcomes hold across a faster, 
conservative judge and a stronger reasoning one. The judge scores each conversation 
against a rubric of twelve metrics across four categories. Goal achievement metrics 
(CFA, ARGA) are bounded in $[0, 1]$ with 1 as the ideal, while the Modality Robustness 
Score targets exactly 1.0, with values below 0.7 indicating the agent is not 
voice-ready. Efficiency metrics include Turn Efficiency ($[0, 1]$, best at 1) and Turn 
Overhead (range $[-1, \infty)$, best at 0), with User Effort Score also minimized 
toward 0. Recovery metrics (ERR in $[0, 1]$, best at 1; RTC in $[0, \infty)$, best 
at 0) capture how well agents handle errors, with domain targets of RTC $\leq 2$ for 
Telecom and Retail. Clarification and safety metrics (Precision and Recall) are 
bounded in $[0, 1]$ with 1 as ideal, where low precision signals over-clarification 
and low recall signals unsafe or ambiguous actions proceeding without confirmation. We 
run all evaluations with and without persona injection to measure how simulated user 
behavior shifts agent performance. The results are reported in 
Table~\ref{tab:combined-results}.

\begin{figure*}[!ht]
    \centering
        \includegraphics[width=0.45\linewidth]{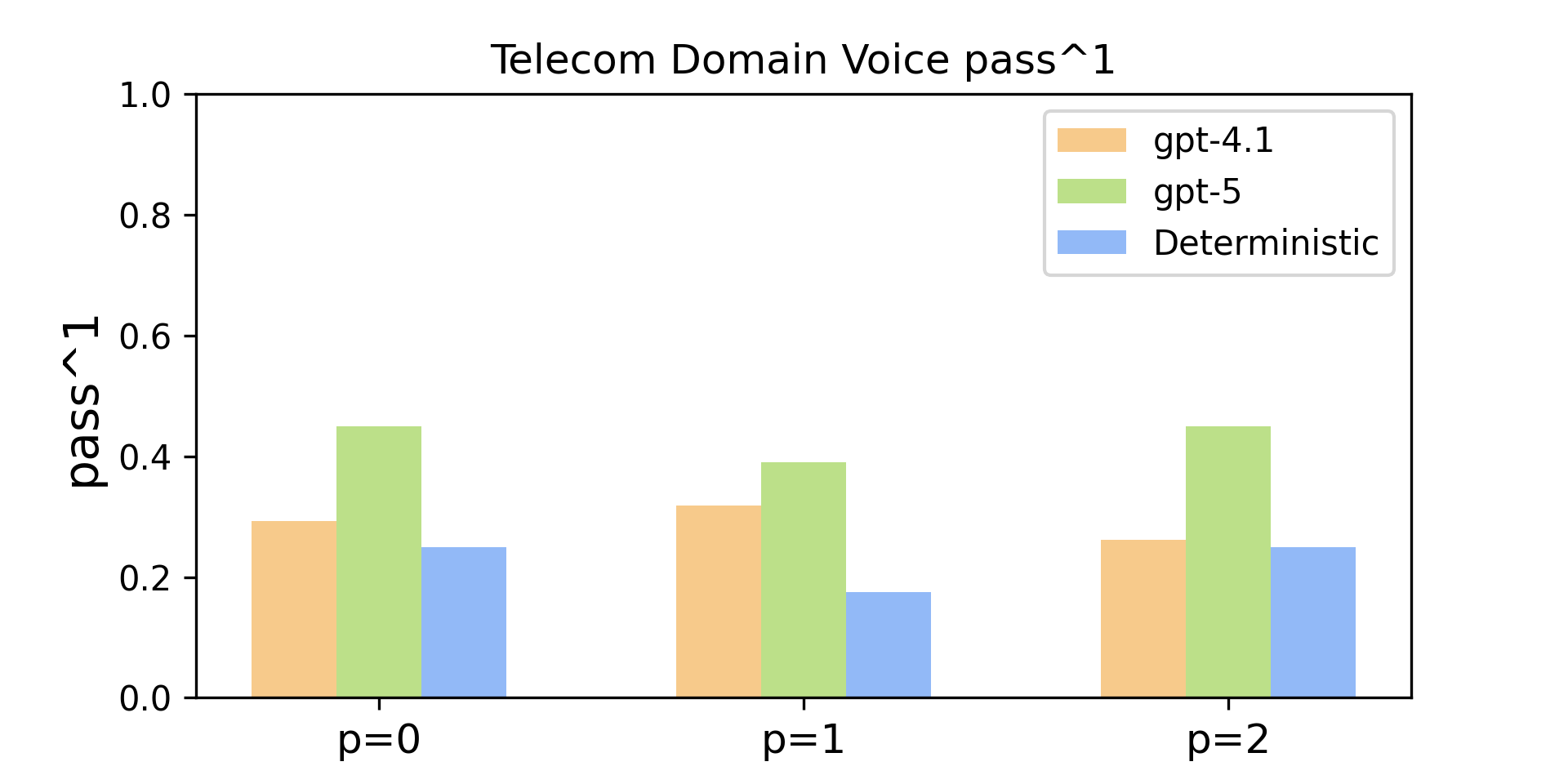}
        \includegraphics[width=0.45\linewidth]{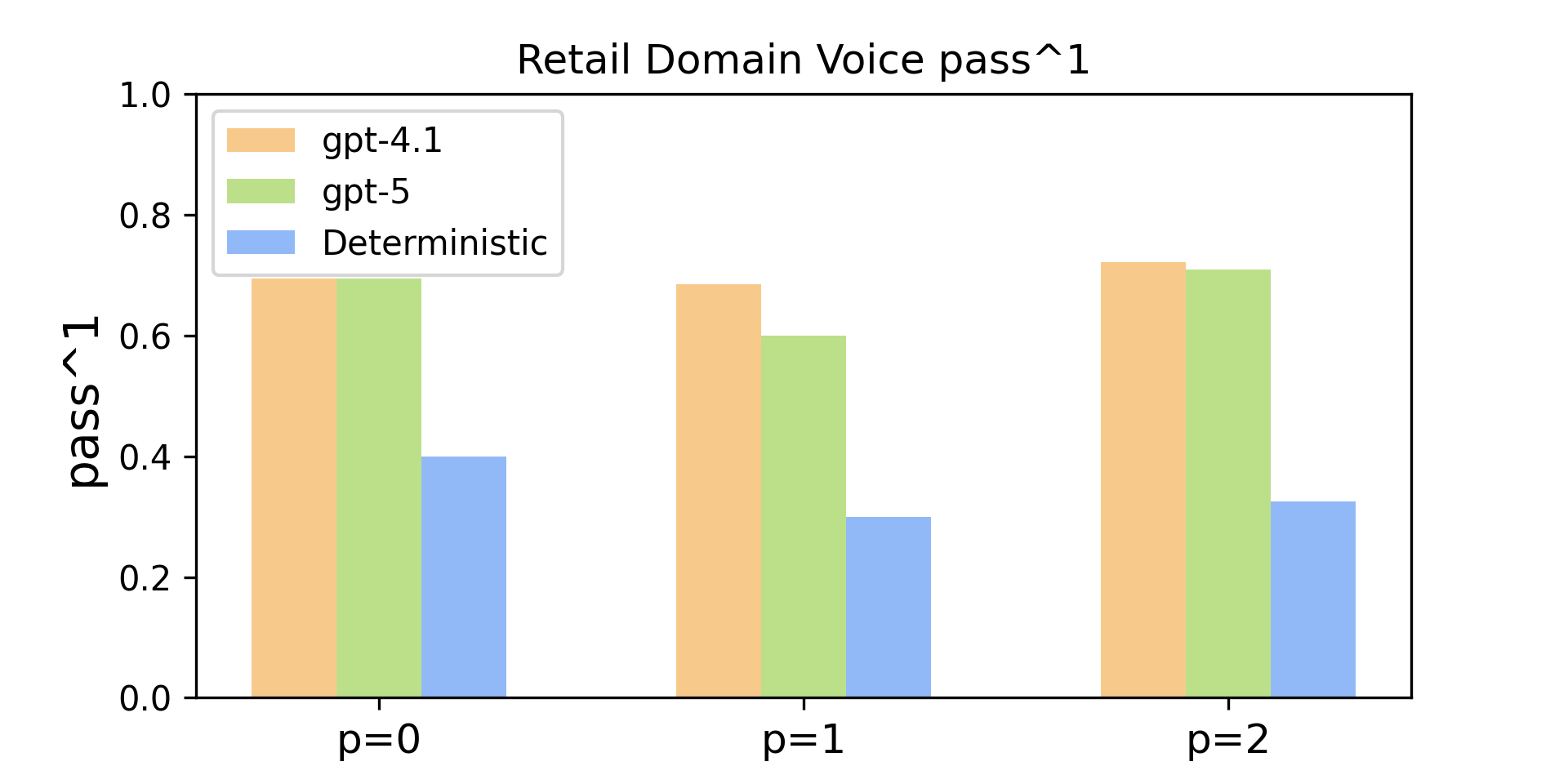}
        \includegraphics[width=0.45\linewidth]{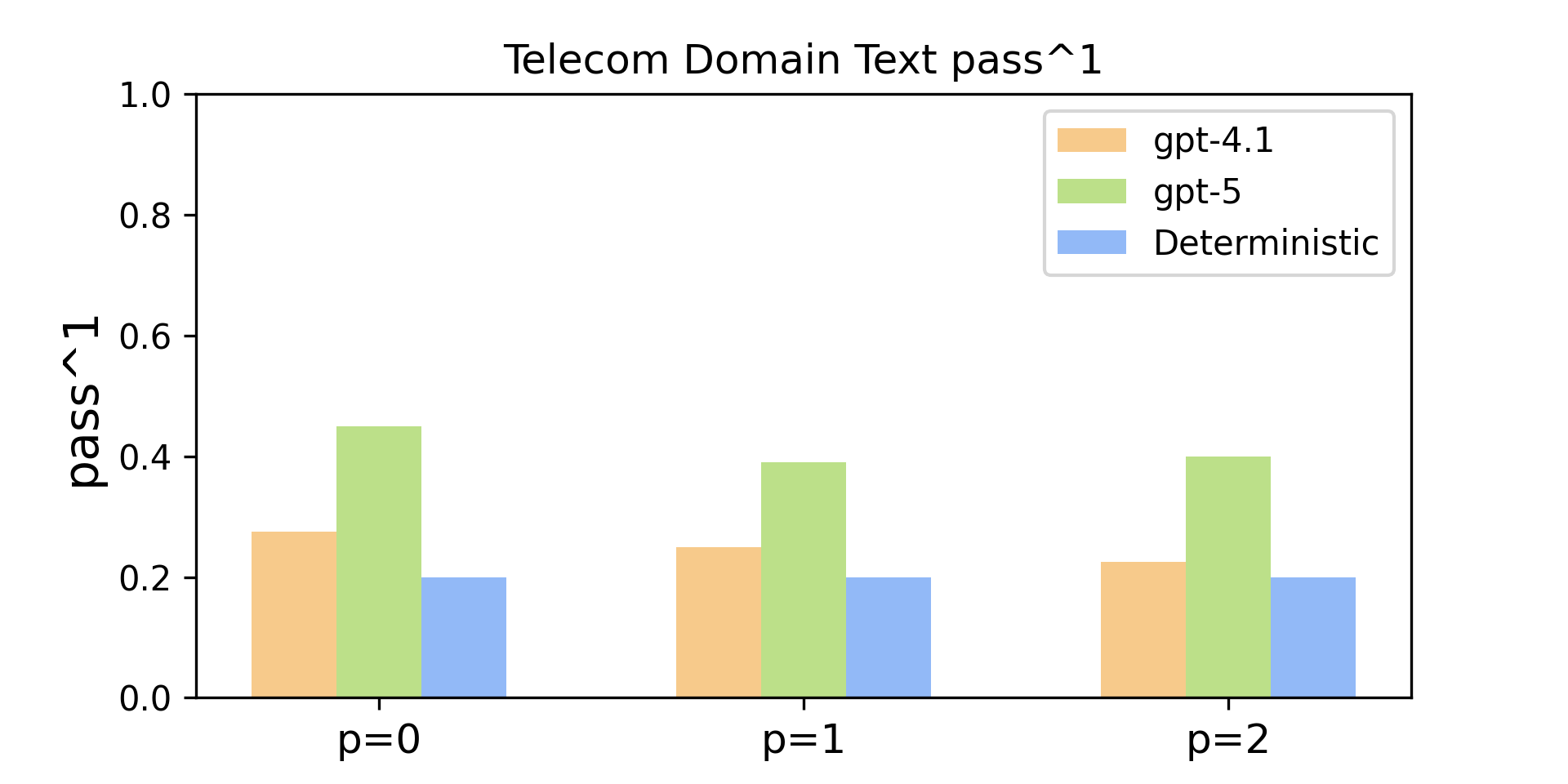}
        \includegraphics[width=0.45\linewidth]{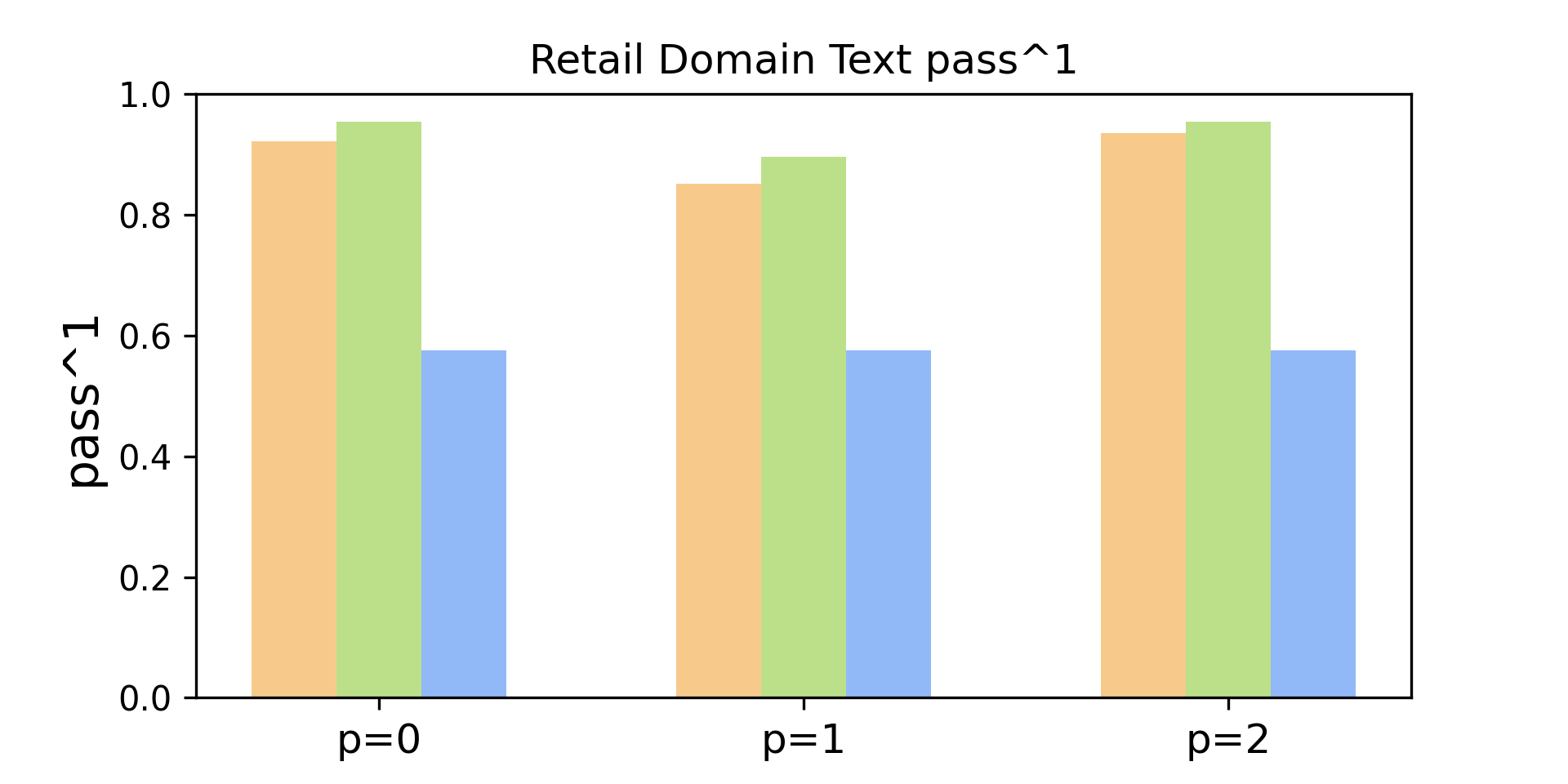}
    \caption{pass\text{\textasciicircum}1 scores for Telecom and Retail domains in Text and Voice Modality. p=0 corresponds to No Persona, p=1 corresponds to Persona Injection and p=2 corresponds to Context Injection}
    \label{fig:pass-retail-telecom}
\end{figure*}

\subsection{GPT-4.1 vs GPT-5 as Judge}
The critical field accuracy is slightly higher under GPT-5 than under GPT-4.1 in both domains. ARGA scores are higher without persona under GPT-5, but persona injection causes a steeper drop than seen under GPT-4.1. Clarification recall is near-perfect under both judges, while clarification precision stays low, indicating that agents over-clarify when context is sufficient. Safety precision and recall are consistently low across all conditions and both judges, marking safety boundary detection as a persistent weak point.

\subsection{Deterministic Evaluation}
The tasks in Tau$^2$-bench are accompanied with an ideal set of actions against which tool-calls by the agent may be compared to identify successful resolution of the task. The list of tool-calls presents the journey of an ideal conversation between the user and the agent.

In order to have a deterministic evaluation for success of tasks, we follow the methodology used in Tau$^2$ codebase and check whether all of the prescribed tool-calls in the task description are used by the agent or not. The Tau$^2$ codebase also performs a state check on user database at the end of the conversation. However we believe that the same was redundant since any and all database modifications happen due to tool-calls. This ensures that the appropriate changes in the database would be found if all the tool-calls are matching with the prescribed list

A database check may alleviate concerns of the agent hallucinating tool call responses. In our approach, the list of tool-calls requested are fetched from info logs of the simulations (as and when the agent requests the environment for executing tool-calls). Thereby, it is ensured that all the tool calls we record are grounded in reality

The approach utilized by the original work has a few faults which may prove detrimental towards accuracy considering the variability present in probabilistic systems such as LLMs.

\begin{itemize}
    \item \textbf{Order of Tool-Calls}: Although The tool calls are listed in the ideal order of calling, the evaluator for the same only checks for the presence of the tool-name and the corresponding rather than the order in which the tools were called. This presents a problem in the cases where order of tool-calls is detrimental to success of a task
    \item \textbf{Statically typed arguments}: The tool-calls are accompanied with the arguments. A statically typed list of arguments presents a problem in certain cases, for example, the retail domain of Tau$^2$ requires a \verb|caculate| tool which takes in an expression and returns the value of the expression. The argument for this tool are provided as a string which is interpreted by the function. However this presents a problem when the data has "1+2" as the argument but in a simulation the agent passes "2+1" to the tool. This leads in the check for tool-calls being a failure due to difference in arguments whereas both the arguments are logically correct.
    \item \textbf{Variance in Conversation Journeys}: The customer intents present in tasks may drive the conversation journey in a direction different to what was intended by the authors. For example, in one of the tasks in retail domain, the user is supposed to inquire about exchanging product A and product B and at the last minute change their mind to acting only on product B. The ideal journey for this requires the agent to perform a tool-call to inquire about product A as well as B. However, in a simulation we ran, the user simulator changes its mind about product A soon after the user's orders are listed. Hence the agent doesn't make any tool-call to inquire about product A. This leads to the conversation being marked a failure since the tool-call for product A was not present. This is a fault at the evaluator's part because the conversation journey took a path where the pre-prescribed tool-call was not necessary.
    
\end{itemize}

To address the shortcomings presented by the deterministic approach used in the original work, and the uncertainty of LLM-as-a-Judge, we propose a hybrid approach wherein LLM-as-a-Judge is provided with the prescribed list of tool calls and the tool-calls executed. An LLM-as-a-Judge would decide on the necessary tool-calls from the prescribed list and then match it against the tool-calls recorded. Comparisons by LLM ensures that the matching is not strict and rather semantic based. The subjectivity in analysis by LLM-as-a-Judge is also resolved by grounding the results in an ideal list of tool-calls.

\subsection{Effect of Detailed Persona Injection}

Table~\ref{tab:retail-pct-change-results} presents evaluation metrics for the Retail domain under three conditions: no persona, persona injection, and context-enriched persona, using both GPT-4.1 and GPT-5 as LLM judges. The Retail domain responds differently to persona injection compared to Telecom. Critical Field Accuracy remains largely stable across all three conditions under both judges. This suggests that retail tasks are less sensitive to persona-driven variability, possibly because the interactions are simpler and more constrained. The ARGA Score is higher in Retail than Telecom overall, but drops sharply under GPT-5 with persona injection, falling from 0.48 to 0.30. This points to judge-sensitive evaluation of goal completion in this domain. With persona injection Turn Overhead shows a negative value, implying that the agent occasionally completes tasks in fewer turns than the reference. This likely reflects over-simplified resolution paths rather than genuine efficiency gains. Recovery Turn Count increases substantially with persona injection and reaches 3.0 under GPT-5, showing that persona conditioning introduces conversational complexity the agent struggles to manage. Modality Robustness stays below 1.0 across all Retail conditions and both judges. With context injection safety metrics show a mild improvement as assessed by GPT-4.1 compared to Telecom, but absolute Safety Precision and Recall values remain far from ideal across all conditioning levels.

Table~\ref{tab:tel-pct-change-results} presents the same analysis for the Telecom domain. Context injection achieves the highest Critical Field Accuracy under both judges, reaching 0.477 for GPT-4.1 and 0.483 for GPT-5. Turn Efficiency, Turn Overhead, and User Effort Score all improve with context enrichment, confirming that detailed context leads to more focused conversations with fewer redundant turns. Modality Robustness stays above 1.0 in the context condition under both judges. Under GPT-5, basic persona injection drops Modality Robustness to 0.87, which is close to the 0.7 voice-readiness threshold. Context enrichment recovers it to 1.125, suggesting that a detailed persona can offset the compounding negative effects of persona and voice modality together. The ARGA Score declines steadily with increasing conditioning under both judges. This reveals a gap between field-level accuracy and goal completion that grows as conditioning becomes richer. The agent gets better at capturing individual task fields but does not assemble them into fully completed outcomes. Safety Precision and Safety Recall decline steadily across all three conditions under both judges. This pattern holds for both GPT-4.1 and GPT-5, making it a judge-invariant signal of genuine behavioral degradation rather than an evaluation artifact. Under GPT-4.1, Safety Recall falls from 0.26 with no persona to 0.133 with context enrichment. This means the agent proceeds without confirmation on potentially unsafe actions in over 83\% of cases at the highest conditioning level, which is a serious reliability concern for a production agent handling account-level operations.

\begin{table*}[!ht]
    \centering
    \small
    \begin{tabular}{|l|cc|cc|cc|cc|cc|cc|}
    \hline
        & \multicolumn{6}{c|}{\textbf{GPT-4.1 Judge}} & \multicolumn{6}{c|}{\textbf{GPT-5 Judge}} \\ \cline{2-13}
        \textbf{Metric} & \multicolumn{2}{c|}{No Persona} & \multicolumn{2}{c|}{Persona} & \multicolumn{2}{c|}{Context} & \multicolumn{2}{c|}{No Persona} & \multicolumn{2}{c|}{Persona} & \multicolumn{2}{c|}{Context} \\ \cline{2-13}
        & Tel & Ret & Tel & Ret & Tel & Ret & Tel & Ret & Tel & Ret & Tel & Ret \\ \hline
        Critical Field Accuracy & 0.47 & 0.38 & 0.49 & 0.38 & 0.48 & 0.39 & 0.48 & 0.44 & 0.49 & 0.44 & 0.48 & 0.43 \\ \hline
        Turn Efficiency & 0.97 & 0.98 & 0.96 & 0.98 & 0.96 & 0.99 & 0.99 & 0.98 & 0.98 & 0.99 & 0.98 & 0.99 \\ \hline
        Turn Overhead & -0.06 & 0.03 & 0.02 & -0.02 & 0.04 & 0.15 & -0.06 & 0.03 & 0.02 & -0.02 & 0.04 & 0.15 \\ \hline
        Error Recovery Rate & 0.25 & 0.07 & 0.18 & 0.05 & 0.36 & 0.04 & 0.16 & 0.09 & 0.10 & 0.05 & 0.15 & 0.09 \\ \hline
        Recovery Turn Count & 0.60 & 0.89 & 0.22 & 2.17 & 0.43 & 2.57 & 1.22 & 1.75 & 0.64 & 3.00 & 0.92 & 1.30 \\ \hline
        User Effort Score & 0.44 & 0.46 & 0.41 & 0.47 & 0.46 & 0.51 & 0.43 & 0.43 & 0.41 & 0.43 & 0.45 & 0.48 \\ \hline
        ARGA Score & 0.32 & 0.51 & 0.23 & 0.44 & 0.26 & 0.58 & 0.35 & 0.48 & 0.28 & 0.30 & 0.32 & 0.59 \\ \hline
        Modality Robustness & 1.04 & 0.75 & 0.94 & 0.80 & 0.86 & 0.77 & 1.03 & 0.73 & 0.87 & 0.67 & 0.88 & 0.74 \\ \hline
        Clarification Precision & 0.17 & 0.68 & 0.12 & 0.70 & 0.17 & 0.97 & 0.17 & 0.47 & 0.11 & 0.43 & 0.15 & 0.25 \\ \hline
        Clarification Recall & 0.96 & 0.98 & 0.89 & 0.92 & 0.94 & 1.00 & 0.98 & 0.97 & 1.00 & 1.00 & 1.00 & 1.00 \\ \hline
        Safety Precision & 0.18 & 0.47 & 0.17 & 0.54 & 0.25 & 0.52 & 0.16 & 0.38 & 0.16 & 0.49 & 0.23 & 0.43 \\ \hline
        Safety Recall & 0.17 & 0.43 & 0.16 & 0.49 & 0.23 & 0.49 & 0.18 & 0.36 & 0.17 & 0.43 & 0.27 & 0.41 \\ \hline
    \end{tabular}
    \caption{Evaluation metrics across Telecom (Tel) and Retail (Ret) domains, with and without persona injection, using GPT-4.1 and GPT-5 as LLM judges. Modality Robustness exceeds 1.0 in Telecom (GPT-4.1, no persona), indicating comparable or better voice performance. GPT-5 yields higher critical field accuracy than GPT-4.1 in both domains, but persona injection degrades ARGA scores more steeply under GPT-5.}
    \label{tab:combined-results}
\end{table*}

\begin{table*}[!ht]
    \centering
    \small
    \setlength{\tabcolsep}{4pt}
    \begin{tabular}{|l|c|c|c|c|c|c|c|}
    \hline
    \multirow{2}{*}{\textbf{Metric}}
        & \multicolumn{3}{c|}{\textbf{GPT-4.1 Judge (Retail)}}
        & \multicolumn{3}{c|}{\textbf{GPT-5 Judge (Retail)}}
        & \multirow{2}{*}{\textbf{Ideal}} \\
    \cline{2-7}
        & \textbf{Context}
        & \textbf{No Persona \%$\Delta$}
        & \textbf{Persona \%$\Delta$}
        & \textbf{Context}
        & \textbf{No Persona \%$\Delta$}
        & \textbf{Persona \%$\Delta$}
        & \\ \hline
    Critical Field Accuracy
        & 0.389 & $-2.3\%$   & $-2.3\%$
        & 0.429 & $+2.6\%$   & $+2.6\%$   & 1.0 \\ \hline
    ARGA Score
        & 0.581 & $-12.2\%$  & $-24.3\%$
        & 0.588 & $-18.4\%$  & $-49.0\%$  & 1.0 \\ \hline
    Modality Robustness
        & 0.772 & $-2.8\%$   & $+3.6\%$
        & 0.743 & $-1.7\%$   & $-9.8\%$   & $\geq$1.0 \\ \hline
    Turn Efficiency
        & \textbf{0.986} & $-0.6\%$   & $-0.6\%$
        & \textbf{0.986} & $-0.6\%$   & $+0.4\%$   & 1.0 \\ \hline
    Turn Overhead
        & 0.146 & $-79.5\%$  & $-113.7\%$
        & 0.146 & $-79.5\%$  & $-113.7\%$ & 0.0 \\ \hline
    User Effort Score
        & 0.507 & $-9.3\%$   & $-7.3\%$
        & 0.479 & $-10.2\%$  & $-10.2\%$  & 0.0 \\ \hline
    Error Recovery Rate
        & 0.041 & $+70.7\%$  & $+22.0\%$
        & 0.085 & $+5.9\%$   & $-41.2\%$  & 1.0 \\ \hline
    Recovery Turn Count
        & 2.570 & $-65.4\%$  & $-15.6\%$
        & 1.300 & $+34.6\%$  & $+130.8\%$ & 0.0 \\ \hline
    Clarification Precision
        & \textbf{0.970} & $-29.9\%$  & $-27.8\%$
        & 0.247 & $+90.3\%$  & $+74.1\%$  & 1.0 \\ \hline
    Clarification Recall
        & \textbf{1.000} & $-2.0\%$   & $-8.0\%$
        & \textbf{1.000} & $-3.0\%$   & $0.0\%$    & 1.0 \\ \hline
    Safety Precision
        & 0.521 & $-9.8\%$   & $+3.6\%$
        & 0.430 & $-11.6\%$  & $+14.0\%$  & 1.0 \\ \hline
    Safety Recall
        & 0.490 & $-12.2\%$  & $0.0\%$
        & 0.408 & $-11.8\%$  & $+5.4\%$   & 1.0 \\ \hline
    \end{tabular}
    \caption{Evaluation metrics for the Retail domain. \textbf{Context} columns show absolute scores under context-enriched persona conditioning. \textbf{\%$\Delta$} columns show the relative change of No Persona and Persona conditions with respect to Context: for higher-is-better metrics (CFA, ARGA, Turn Efficiency, Error Recovery Rate, Clarification and Safety Precision/Recall), negative values indicate underperformance relative to Context; for lower-is-better metrics (Turn Overhead, User Effort Score, Recovery Turn Count), positive values indicate underperformance. The \textbf{Ideal} column denotes the theoretical optimum. Bold values indicate the best result per metric per judge.}
    \label{tab:retail-pct-change-results}
\end{table*}


\begin{table*}[!ht]
    \centering
    \small
    \setlength{\tabcolsep}{4pt}
    \begin{tabular}{|l|c|c|c|c|c|c|c|}
    \hline
    \multirow{2}{*}{\textbf{Metric}}
        & \multicolumn{3}{c|}{\textbf{GPT-4.1 Judge (Telecom)}}
        & \multicolumn{3}{c|}{\textbf{GPT-5 Judge (Telecom)}}
        & \multirow{2}{*}{\textbf{Ideal}} \\
    \cline{2-7}
        & \textbf{Context}
        & \textbf{No Persona \%$\Delta$}
        & \textbf{Persona \%$\Delta$}
        & \textbf{Context}
        & \textbf{No Persona \%$\Delta$}
        & \textbf{Persona \%$\Delta$}
        & \\ \hline
    Critical Field Accuracy
        & 0.478 & $-0.65\%$   & $+1.82\%$   
        & 0.482 & $+0.51\%$   & $+1.42\%$   & 1.0 \\ \hline
    ARGA Score
        & 0.261          & $+22.63\%$  & $-12.55.0\%$  
        & 0.315          & $+11.46\%$  & $-10.64\%$   & 1.0 \\ \hline
    Modality Robustness
        & 0.864          & $+20.38\%$   & $+8.25\%$   
        & 0.879 & $+16.89\%$  & $-0.75\%$  & $\geq$1.0 \\ \hline
    Turn Efficiency
        & 0.96 & $+0.66\%$   & $+0.26\%$   
        & 0.982 & $+0.66\%$   & $+0.2\%$   & 1.0 \\ \hline
    Turn Overhead
        & 0.038 & $-266.37\%$  & $-38.37\%$ 
        & 0.038 & $-266.37\%$  & $-38.37\%$ & 0.0 \\ \hline
    User Effort Score
        & 0.461 & $-3.5\%$  & $-10.13\%$   
        & 0.445 & $-3.98\%$   & $-7.25\%$   & 0.0 \\ \hline
    Error Recovery Rate
        & \textbf{0.358}          & $-30.99\%$  & $-49.01\%$  
        & 0.152          & $+5.15\%$  & $-35.09\%$   & 1.0 \\ \hline
    Recovery Turn Count
        & 0.431 & $+38.1\%$ & $-49.57\%$  
        & 0.917          & $+32.81\%$   & $-29.87\%$   & 0.0 \\ \hline
    Clarification Precision
        & \textbf{0.169}          & $-1.84\%$  & $-29.2\%$  
        & 0.147          & $+16.18\%$  & $-23.53\%$  & 1.0 \\ \hline
    Clarification Recall
        & 0.944          & $+1.76\%$   & $-5.88\%$   
        & 1.00          & $-1.61\%$   & $+0.0\%$   & 1.0 \\ \hline
    Safety Precision
        & \textbf{0.248}          & $-28.28\%$  & $-31.24\%$  
        & \textbf{0.233}          & $-30.98\%$  & $-29.56\%$  & 1.0 \\ \hline
    Safety Recall
        & \textbf{0.235}          & $-29.37\%$  & $-31.31\%$  
        & \textbf{0.266}          & $-33.89\%$  & $-35.85\%$  & 1.0 \\ \hline
    \end{tabular}
    \caption{Evaluation metrics for the Telecom domain. \textbf{Context} columns show absolute scores under context-enriched persona conditioning. \textbf{\%$\Delta$} columns show the relative change of No Persona and Persona conditions with respect to Context: for higher-is-better metrics (CFA, ARGA, etc.), negative values mean those conditions underperformed Context; for lower-is-better metrics (Turn Overhead, User Effort Score, Recovery Turn Count), positive values indicate underperformance. The \textbf{Ideal} column denotes the theoretical optimum.}
    \label{tab:tel-pct-change-results}
\end{table*}

\begin{figure*}[!ht]
    \centering
        \includegraphics[width=0.45\linewidth]{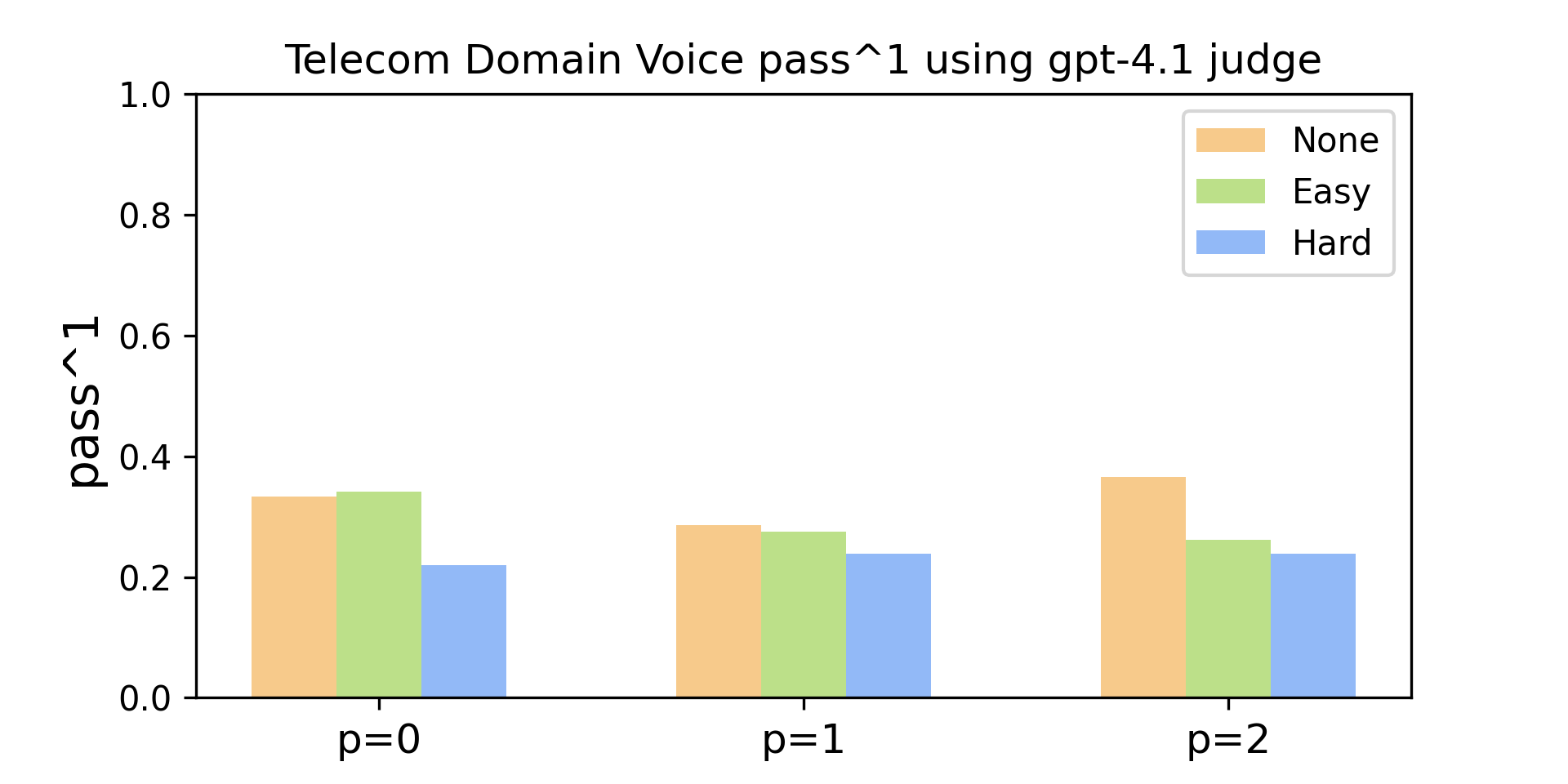}
        \includegraphics[width=0.45\linewidth]{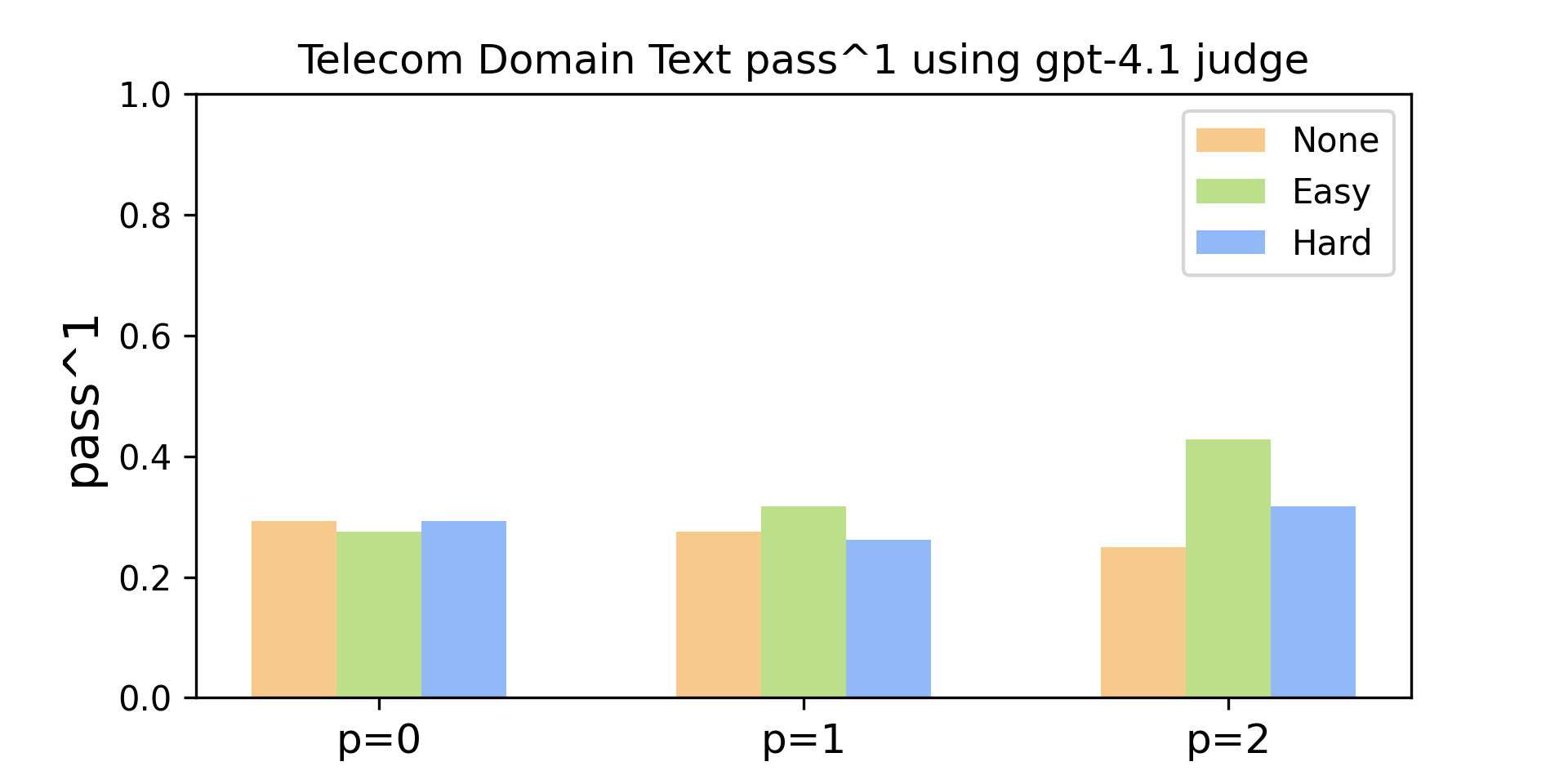}
        \includegraphics[width=0.45\linewidth]{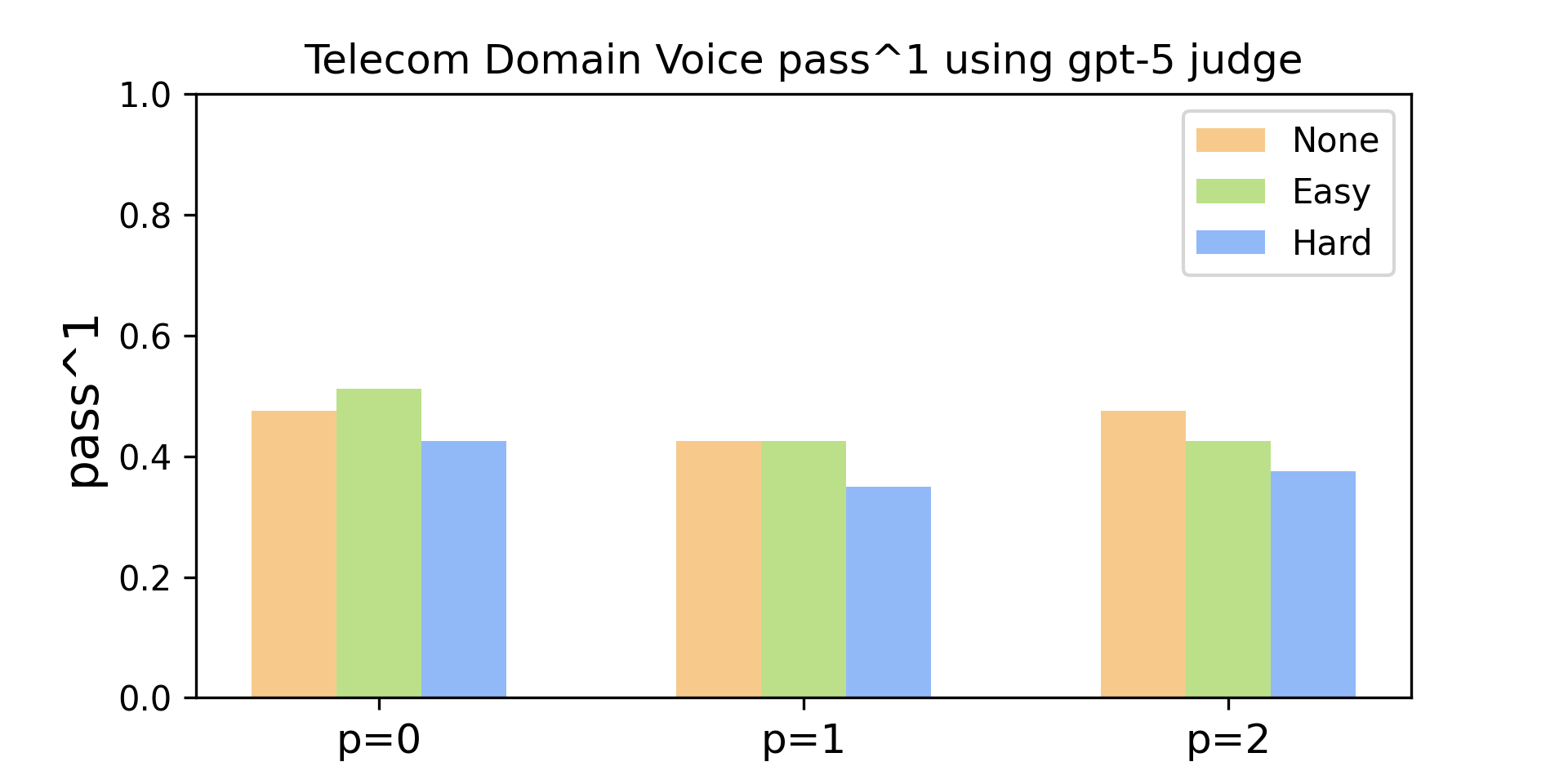}
        \includegraphics[width=0.45\linewidth]{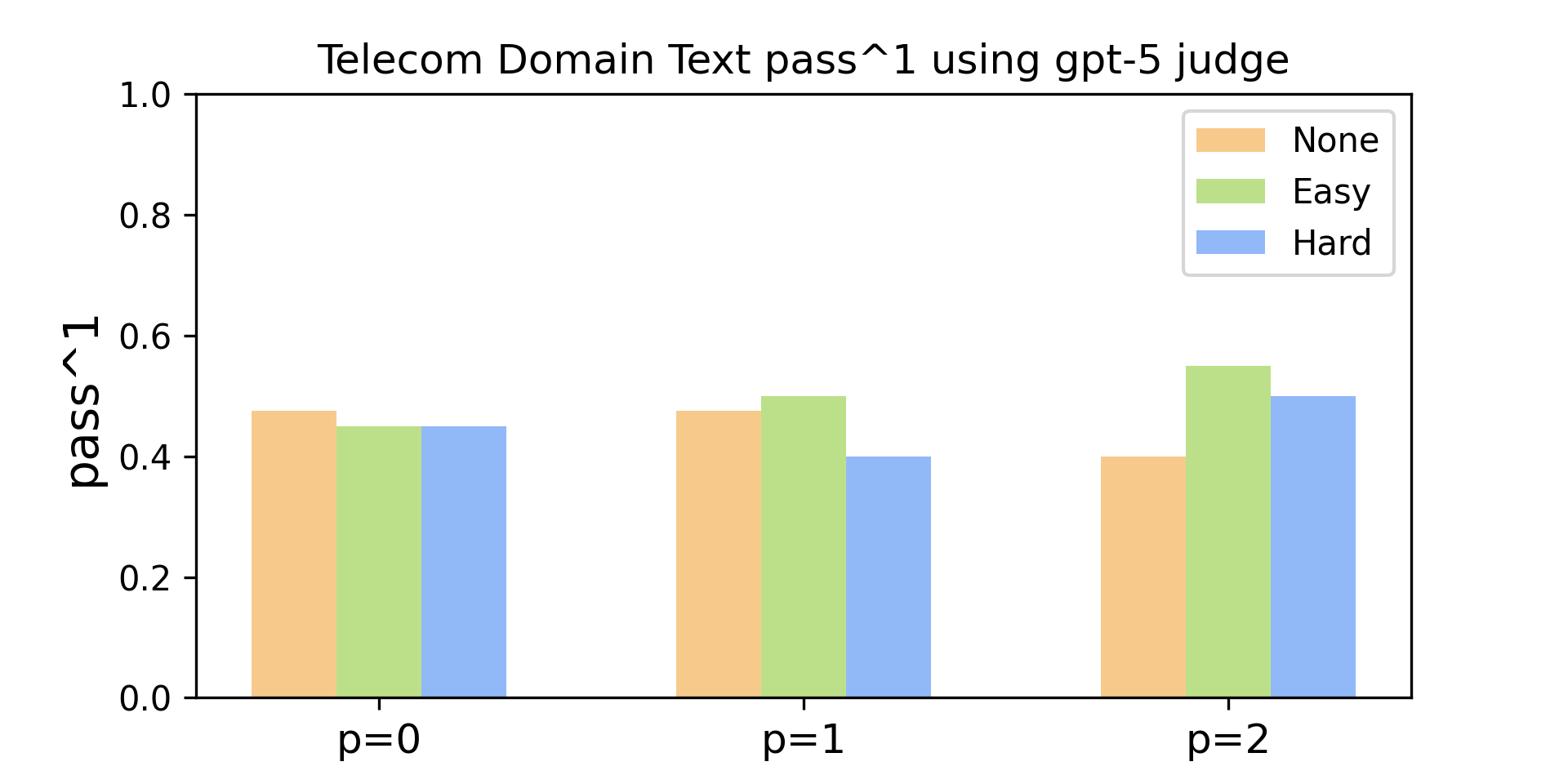}
    \caption{pass$^1$ scores segregated by persona difficulty in Telecom domain in Text and Voice Modality. p=0 corresponds to No Persona, p=1 corresponds to Persona Injection and p=2 corresponds to Context Injection}
    \label{fig:persona}
\end{figure*}

\subsection{Optimal value of p}
Analyzing the success of various $p$ modes portrays a mixed picture as presented in Fig. \ref{fig:persona}. For tasks having a user well versed with the domain (Easy Persona), the text modality prefers $p=0$ and voice modality prefers $p=1$. For tasks having no defined user persona (None), we witness the flipped scenario with text modality preferring $p=1$ and the voice modality preferring $p=0$. In both these cases, both gpt-4.1 and gpt-5 judge agree on the optimal value of $p$ chosen. 
A simple persona injection and $p=0$ fare better as they are more focused on interacting with the user coupled with the fact that user is able to contribute to the diagnosis. Context injection rather focuses on generating dynamic responses which is not required, as the user is mature enough to not require extra care from the agent.

In case of tasks involving a user not well versed with the domain (Hard), a simple persona injection doesn't aid the performance and rather degrades it across all modalities and judges. This can be attributed to the fact that the special assistance that the user requires changes dynamically through the conversation.
Persona injection models only a static persona for the user which prevents the agent to adapt to the needs of the user at each step. No persona/context injection and context injection on the other hand are able to able to dynamically adapt to responding to the user.

\section{Discussions}
MM-tau-p$^2$ reveals consistent, domain-dependent differences in agent behavior under multi-modal interaction. In the Retail domain, tasks tend to remain well structured, with fewer conversational detours and fewer instances where the agent abandons resolution in favor of escalation. In the Telecom domain, we observe qualitatively higher conversational fragility: the agent more often shifts toward escalation-like behavior when faced with ambiguity, missing critical fields, or long repair loops. This is owing to higher complexity of telecom task, refer Appendix \ref{tdt}, which is also evident from lower pass$^1$ scores for telecom domain, refer Fig. \ref{fig:pass-retail-telecom}. These tendencies are amplified when voice is introduced due to transcript noise and additional clarification turns. Across both domains, the multi-modal setting increases interaction cost through additional repair and confirmation turns, which is reflected in higher turn overhead and lower modality robustness relative to text-only runs. These observations support the need for evaluation that jointly measures task success, safety, and efficiency under modality shifts, rather than relying on a single pass rate.

\subsection{LLM as Judge}
MM-tau-p$^2$ relies on an LLM-as-judge pipeline to score every conversation automatically. The judge receives a rubric-augmented prompt containing metric definitions (e.g., \texttt{tasks\_identified}, \texttt{tasks\_succeeded}) and is asked to return a structured evaluation for each conversation. These per-conversation counts feed directly into aggregate metrics. Although the judge was prompted to return a JSON schema with boolean/integer fields for each sub-criterion, rather than a free-form narrative followed by a score we observed different conclusions across runs. \\
To resolve the escalation ambiguity, we iteratively refined the judge prompt Appendix Table \ref{tab:misclassify}. The original instruction stated: \textit{``A task is considered resolved when the agent provides a satisfactory resolution and transfers to a human agent only if absolutely necessary.''}  We updated it with a domain-specific details for manadatory escalation to human agent (in case of Telecom domain, issues related to locked SIM need to be escalated). This update reduced but did not eliminate the inconsistency: the judge still produced split verdicts in 3 SIM-lock escalation cases while 2 of the previous cases were resolved. Table Appendix \ref{sec:misclassify} documents the iterative process followed.

\subsubsection{Inconsistent treatment of escalations}
The most prevalent failure we observe is the mis-classification of \textit{tasks\_succeeded} when the agent correctly identifies that a user's problem and escalates to a human agent, an action though correct but is marked as failure. In the Telecom domain, a common diagnosis is a SIM PIN lock as the root cause and the transfers the user to a human agent as it requires actions that agent cannot perform. This instance is a correct escalation and hence should have been counted as a successful task. When presented with this scenario, both GPT-4.1 and GPT-5 judges produce contradictory labels across runs on similar conversations under identical prompts. In one evaluation, the judge reasons that \textit{"the agent successfully resolved the task to the best of its ability, identifying the root cause (SIM PIN lock) and appropriately transferring to a human agent"} and marks the task as succeeded. On the other hand, in a parallel evaluation the judge reasons that \textit{"the task was not fully resolved by the agent \ldots required escalation to a human agent \ldots so the task is not counted as succeeded."} Instances of this mis-classification are documented in Appendix \ref{sec:misclassify}. The dichotomy arises from an ambiguous boundary between agent failure and escalation due to limited scope, a distinction the rubric must encode but that natural-language instructions alone do not resolve unambiguously.

Evaluations show cases where the judge acknowledge that \textit{the agent followed a structured troubleshooting process, checking network mode, APN settings, permissions, and SIM card reseating}, yet it concludes that \textit{the task was not successfully resolved} solely because a human transfer occurred. Meanwhile, in structurally similar conversations, the judge credits agent's efforts as sufficient for success. This discrepancy has major impact on $\text{pass}^k$ because it introduces label noise that is correlated with task difficulty. Harder tasks are more likely to require escalation, so the judge's inconsistency disproportionately affects harder tasks.

The Telecom domain by design has harder tasks as compared to the Retail domain. Hence, Modality Robustness Score (MRS) witnesses a substantial change between the two. The Telecom domain is more sensitive to faults of the LLM judge as the text and the voice agent have similar performance in most instances, unlike Retail domain where the two diverge often in the verification step.
In the Retail domain, verification has lower success rate for the Voice based agent owing to the fact that ASR transcriptions may not accurately capture nuances of spellings in names and e-mail addresses. This leads to more frequent failures in user verification as compared to the Telecom domain.

\subsubsection{LLM Judge: gpt-4.1 vs. gpt-5}

We evaluated the conversations using \verb|gpt-5-chat-latest| and \verb|gpt-4.1| as judges. GPT-5 took an optimistic outlook by classifying most tasks that escalated to a human agent as success. Its reasoning was that the agent had tried everything it could, and therefore escalation was justified. While this correctly handled valid escalations that would otherwise be marked as failure, it also introduced false positives where escalations that should have been marked as failure were counted as success. This led to a substantial increase in pass$^{1}$  scores for both text and voice agents across domains.
GPT-5 consistently assigned higher pass rates than GPT-4.1 across nearly all domain and modality conditions, refer \ref{fig:pass-retail-telecom}. The gap was most pronounced in the Telecom domain, particularly for voice interactions, where GPT-5 scores exceeded GPT-4.1 by up to 17 percentage points. In the Retail domain the two judges were more closely aligned, likely because the tasks are clearer and less affected by acoustic noise from voice input. Overall, these differences raise questions about calibration consistency when using large language models as judges, and suggest that the choice of judge model can meaningfully influence evaluation outcomes.

\subsection{Challenges}
Evaluating multi-modal (voice/text) agents with a human simulator introduces structural challenges that do not arise in text-only settings.

\begin{itemize}[itemsep=-0.4pt]
    \item \textbf{Pipeline mismatch.} Voice requires an ASR $\rightarrow$ LLM $\rightarrow$ TTS sequence, while text does not. This means the two modalities follow different inference paths, and even identical prompts can produce different outputs. ASR artifacts, prosody shifts, or channel noise can subtly change what the agent hears, which can cascade into diverging conversations including premature task resolution or unintended escalation. To manage this, the simulator requires additional guardrail prompts.
    \item \textbf{Conversational rhythm.} Multi-modal agents must meet stricter requirements around turn-taking and latency. These are far more sensitive in voice compared to text interactions.
    \item \textbf{Measurement inconsistency.} The LLM agent may produce different responses for identical inputs depending on whether the trigger is voice or text. This makes reliable, controlled measurement difficult.
    \item \textbf{Simulator variability.} Because the human simulator reacts to agent outputs, any agent-level variation propagates into the simulator's responses as well. Voice and text interactions therefore cannot be treated as like-for-like comparisons.
\end{itemize}
 
\begin{figure*}[t]  
  \begin{equation}
    \begin{split}
      S &= \alpha (\text{CFA} \cdot \text{pass}^1) 
          + \beta\frac{\text{MRS}+\text{ARGA}+\text{ERR}}{3} 
    + \gamma\frac{\text{TE}+\frac{1}{1+\text{UES}}+\frac{1}{1+\text{TO}}}{3} + \delta\,\text{IAS}
    \end{split}
  \end{equation}
  \vspace{-0.5em}
  {\small mm-tap score, where CFA = Critical Field Accuracy; MRS = Modality Robustness Score; 
  ARGA = ASR Robust Goal Achievement; ERR = Error Recovery Rate; 
  TE = Turn Efficiency; TO: Turn Overhead; UES = User Experience Score; IAS = Irreversible Action Safety; 
  $\alpha, \beta, \gamma, \delta$ = configurable weights which add to 1}
\end{figure*}

Effective multi-modal evaluation must separately assess the quality of understanding, reasoning, and response delivery, intent determination under noisy conditions, maintenance of multi-turn context, and tone/prosody control. Besides, measuring the metrics in an automated way requires careful crafting of LLM prompt \cite{tan2025judgebenchbenchmarkevaluatingllmbased} \cite{Wang_2025} \cite{feng2025rightwayassessingllmasajudge}. Taken together, these considerations show that multi-modal evaluation cannot be accomplished by simply inserting TTS and ASR components around a text agent. It instead requires guardrail prompts for the human simulator, tight instructions to LLM agent for uniformity in agent's response. Last but not the least, one needs a composite metric to perform comparisons across architectures of multi-modal agents. Towards this, we propose a composite score viz. \textbf{mm-tap} to rank performance of  multi-modal agents, refer Equation 1.

\begin{table}[!ht]
    \centering
    \small
    \begin{tabular}{|l|ccc|ccc|}
    \hline
        & \multicolumn{3}{c|}{\textbf{Telecom}} & \multicolumn{3}{c|}{\textbf{Retail}} \\
        p & 0 & 1 & 2 & 0 & 1 & 2\\ \hline
        gpt-4.1 & 0.437 & 0.396 & 0.42 & 0.470 & 0.474 & 0.481 \\ 
        gpt-5 & 0.459 & 0.409 & 0.428 & 0.473 & 0.441 & 0.484 \\ \hline
    \end{tabular}
    \caption{\textbf{mm-tap} for gpt-4.1 and gpt-5 judge calculated with parameters $\alpha=0.35$ $\beta=0.35$ $\gamma=0.2$ $\delta=0.1$}
    \label{tab:composite-scores}
\end{table}

\section{Conclusion}
Benchmarking agents with multi-modal inputs has been attempted in this work by introducing novel metrics and a composite agent evaluation score to comprehensively measure an agent's performance. We evaluated the performance of agents using GPT4.1, GPT-5 as LLM backbone while using ElevenLabs as text-to-speech service. Our experiments reveal the limits of LLM-as-judge in evaluating these metrics. Here are the salient findings of our investigation.

\begin{itemize}
    \item MM-tau-p$^2$ introduces 12 novel metrics across four categories covering goal achievement, efficiency, recovery, and safety to evaluate multi-modal agents in dual-control settings.

    \item Context-enriched persona injection improves Critical Field Accuracy and conversational efficiency but consistently degrades safety metrics across both domains and both judges, revealing a trade-off between efficiency and safe task execution.

    \item Safety Precision and Safety Recall decline monotonically with increasing persona conditioning and remain critically low across all settings, indicating that safety boundary detection is a persistent weak point in current frontier LLM agents.

    \item GPT-5 consistently assigns higher pass rates than GPT-4.1, particularly in the Telecom voice setting where scores differ by up to 17 percentage points, showing that the choice of LLM judge meaningfully influences evaluation outcomes.

    \item The LLM-as-judge approach produces inconsistent verdicts on escalation scenarios, where structurally similar conversations receive opposite labels, introducing label noise that is correlated with task difficulty.

    \item For tasks involving users with lower domain expertise, static persona injection degrades performance across all modalities. Context injection adapts dynamically to user behavior and is better suited for handling novice users.

    \item The composite mm-tap score provides a single comparable metric across agent architectures and conditioning levels, enabling holistic ranking of multi-modal agent performance beyond pass rate alone.
\end{itemize}

\section*{Acknowledgement}
Authors acknowledge Yoginkumar Patel and Amitabh Misra for their encouragement to drive innovation through research.

\section*{Future Work}
In the current benchmark, we did not model missed response windows, interruption handling and overtalk which equate with user abandonment. These phenomena do not surface in purely text interaction but are important to voice interactions with agent and can be accounted for addition in future work.

\bibliography{references}

\newpage
\onecolumn
\appendix
\renewcommand{\thesection}{\Alph{section}}
\renewcommand{\thesubsection}{\Alph{section}.\arabic{subsection}}

\section{Appendix}

\subsection{Telecom Domain Tasks}
\label{tdt}
\begin{itemize}

\item[--] [mms\_issue]airplane\_mode\_on | bad\_network\_preference | bad\_wifi\_calling | break\_apn\_mms\_setting | break\_app\_both\_permissions | unseat\_sim\_card | user\_abroad\_roaming\_enabled\_off
\item[--] [mms\_issue]airplane\_mode\_on | bad\_network\_preference | bad\_wifi\_calling | break\_apn\_mms\_setting | break\_app\_sms\_permission | data\_mode\_off | data\_usage\_exceeded | unseat\_sim\_card | user\_abroad\_roaming\_disabled\_on
\item[--] [mms\_issue]airplane\_mode\_on | bad\_network\_preference | bad\_wifi\_calling | break\_apn\_mms\_setting | break\_app\_storage\_permission | data\_mode\_off | data\_usage\_exceeded | unseat\_sim\_card | user\_abroad\_roaming\_disabled\_on
\item[--] [mms\_issue]airplane\_mode\_on | bad\_network\_preference | break\_app\_both\_permissions | data\_usage\_exceeded | unseat\_sim\_card | user\_abroad\_roaming\_disabled\_on
\item[--] [mms\_issue]airplane\_mode\_on | bad\_network\_preference | break\_app\_storage\_permission | data\_mode\_off | user\_abroad\_roaming\_enabled\_off
\item[--] [mms\_issue]airplane\_mode\_on | bad\_wifi\_calling | break\_app\_both\_permissions | data\_mode\_off | data\_usage\_exceeded | unseat\_sim\_card | user\_abroad\_roaming\_enabled\_off
\item[--] [mms\_issue]airplane\_mode\_on | bad\_wifi\_calling | user\_abroad\_roaming\_enabled\_off
\item[--] [mms\_issue]airplane\_mode\_on | break\_app\_both\_permissions
\item[--] [mms\_issue]airplane\_mode\_on | break\_app\_both\_permissions | data\_usage\_exceeded | user\_abroad\_roaming\_disabled\_off
\item[--] [mms\_issue]bad\_network\_preference | bad\_wifi\_calling | break\_app\_sms\_permission | data\_mode\_off | data\_usage\_exceeded | unseat\_sim\_card | user\_abroad\_roaming\_enabled\_off
\item[--] [mms\_issue]bad\_network\_preference | break\_app\_both\_permissions
\item[--] [mms\_issue]bad\_network\_preference | break\_app\_sms\_permission | user\_abroad\_roaming\_disabled\_on
\item[--] [mms\_issue]bad\_network\_preference | data\_mode\_off | user\_abroad\_roaming\_disabled\_on
\item[--] [mms\_issue]bad\_wifi\_calling | break\_apn\_mms\_setting | break\_app\_both\_permissions | data\_mode\_off | data\_usage\_exceeded | user\_abroad\_roaming\_disabled\_off
\item[--] [mms\_issue]break\_apn\_mms\_setting | user\_abroad\_roaming\_enabled\_off
\item[--] [mms\_issue]break\_app\_sms\_permission | data\_mode\_off
\item[--] [mobile\_data\_issue]airplane\_mode\_on | bad\_network\_preference | bad\_vpn | data\_mode\_off | data\_saver\_mode\_on | data\_usage\_exceeded | user\_abroad\_roaming\_enabled\_off
\item[--] [mobile\_data\_issue]airplane\_mode\_on | bad\_network\_preference | data\_mode\_off | data\_saver\_mode\_on
\item[--] [mobile\_data\_issue]airplane\_mode\_on | data\_mode\_off | data\_saver\_mode\_on | data\_usage\_exceeded | user\_abroad\_roaming\_enabled\_off
\item[--] [mobile\_data\_issue]airplane\_mode\_on | data\_saver\_mode\_on | user\_abroad\_roaming\_disabled\_on
\item[--] [mobile\_data\_issue]bad\_network\_preference | bad\_vpn | data\_mode\_off | data\_saver\_mode\_on | data\_usage\_exceeded | user\_abroad\_roaming\_enabled\_off
\item[--] [mobile\_data\_issue]bad\_network\_preference | bad\_vpn | data\_saver\_mode\_on | data\_usage\_exceeded | user\_abroad\_roaming\_disabled\_off
\item[--] [mobile\_data\_issue]bad\_network\_preference | bad\_vpn | user\_abroad\_roaming\_disabled\_off
\item[--] [mobile\_data\_issue]bad\_vpn | data\_mode\_off | data\_usage\_exceeded | user\_abroad\_roaming\_disabled\_off
\item[--] [mobile\_data\_issue]data\_saver\_mode\_on | user\_abroad\_roaming\_enabled\_off
\item[--] [service\_issue]airplane\_mode\_on | break\_apn\_settings | contract\_end\_suspension | lock\_sim\_card\_pin
\item[--] [service\_issue]airplane\_mode\_on | break\_apn\_settings | contract\_end\_suspension | lock\_sim\_card\_pin | unseat\_sim\_card
\item[--] [service\_issue]airplane\_mode\_on | break\_apn\_settings | lock\_sim\_card\_pin | overdue\_bill\_suspension
\item[--] [service\_issue]airplane\_mode\_on | break\_apn\_settings | lock\_sim\_card\_pin | unseat\_sim\_card
\item[--] [service\_issue]airplane\_mode\_on | break\_apn\_settings | overdue\_bill\_suspension
\item[--] [service\_issue]airplane\_mode\_on | contract\_end\_suspension | lock\_sim\_card\_pin | unseat\_sim\_card
\item[--] [service\_issue]airplane\_mode\_on | lock\_sim\_card\_pin
\item[--] [service\_issue]airplane\_mode\_on | lock\_sim\_card\_pin | overdue\_bill\_suspension
\item[--] [service\_issue]airplane\_mode\_on | overdue\_bill\_suspension
\item[--] [service\_issue]break\_apn\_settings | contract\_end\_suspension | lock\_sim\_card\_pin
\item[--] [service\_issue]break\_apn\_settings | contract\_end\_suspension | lock\_sim\_card\_pin | unseat\_sim\_card
\item[--] [service\_issue]break\_apn\_settings | lock\_sim\_card\_pin
\item[--] [service\_issue]break\_apn\_settings | lock\_sim\_card\_pin | overdue\_bill\_suspension
\item[--] [service\_issue]contract\_end\_suspension | unseat\_sim\_card
\item[--] [service\_issue]overdue\_bill\_suspension | unseat\_sim\_card
\end{itemize}

\subsection{Prompt for User context extraction}
\label{pruce}
\begin{minipage}[t]{\textwidth}
\begin{boxD}
\label{context_injection_prompt}
\begin{lstlisting}
You are an expert in determining personality of users.
### TASK
You are given a conversation between a user and an assistant. Analyse the conversations and generate a persona of the user. Make sure to infer the following fields
{
"age": "[int] Infered age of user",
"gender": "[str] Infered gender of the user",
"articulation_skill": "[int] Score user's ability to express their problem and communications clearly on a scale of 10",
"language_understanding_skill": "[int] Score user's ability to understand agent's responses on a scale of 10",
"technical_skill": "[int] Score user's ability to follow along technical steps instructed by agent on a scale of 10",
"frustration_level": "[int] Score the user's frustrated with the process on a scale of 10",
"persona": "[str] Describe user's persona in natural langugage",
<misc metric>: "<appropriate score on a scale of 10>",
"reasoning" : "provide reasoning for the scores given and the persona infered"
}

Within misc metrics, include other fields which would be helpful for the assistant to understand the user and help them or any other personality trait you could infer from the conversation
DO NOT generate a persona which is not grounded in conversation history. If you fail to infer any of the fields above then do not include them in your answer. Output "N/A" in metrics which could not be infered

### Example
[Conversation]
USER: When I try to send an MMS, it just fails to send. There's no specific error message. Also, it looks like I'm not able to use mobile data or make regular calls right now. What should I do next?
ASSISTANT: Your line, 5551232002, appears to be active and your device is properly activated ... If you need guidance on how to check, let me know.
USER: I checked my phone's status bar and I see the airplane mode icon is on. I don't see any signal bars, just the airplane mode symbol. What should I do next?
ASSISTANT: That explains why you're unable to make calls, use mobile data, or send MMS. ... After that, try making a call or sending an MMS. Let me know if this resolves the issue or if you still have trouble.
USER: I've turned off airplane mode, but now my status bar shows no signal and data is still disabled. I still can't make calls or send MMS. What should I try next?

[Output]
{
"age": "N/A",
"gender": "N/A",
"articulation_skill": "8",
"language_understanding_skill": "10",
"technical_skill": "8",
"frustration_level": "0",
"persona": "The user has good articulation skills and the user is able to understand agent's langugage and follow the instructions. The user is not frustrated yet but is eager to resolve their issue",
"eagerness: "8",
"reasoning" : "The age and gender could not be identified. The user has good articulation skills, as the user is able to explain their problem clearly without hesitation and the user is able to understand agent's langugage and follow the instructions precisely demonstrating technical skill. The user is not frustrated yet but is eager to resolve their issue as indicated by them asking "What should I do next?" repeatedly"
}

Your output should be a python dictionary (which can be parsed and evaluated directly) in the format used to describe the metrics. The persona will be passed to the agent for downstream tasks. Evaluate the persona based on following conversation

<CONVERSATION>

BEGIN NOW
\end{lstlisting}
\end{boxD}
\end{minipage}

\subsection{LLM Judge Prompt: Iterative Creation}
\label{sec:misclassify}

\begin{longtable}[htbp]{|p{3cm}|p{1cm}|p{3cm}|p{1.9cm}|p{1.05cm}|p{3.5cm}|}
    \caption{Iterative prompt patching} \\
  \toprule
    \textbf{Example conversation} & \textbf{LLM Judge} & \textbf{Original Prompt} & \textbf{Prompt Ambiguity} & \textbf{Update worked?} & \textbf{Updated prompt}  \\
    \hline
    
    The task was not successfully resolved by the agent; the user had to be transferred to a human agent after all troubleshooting steps failed VS The agent successfully troubleshoots the issue up to the point where a human agent is required for security reasons (SIM PIN/PUK retrieval), which is appropriate and not a failure of the agent & gpt-4.1 & Give the number of tasks successfully resolved. A task is considered resolved when the agent is able to provide a satisfactory resolution to the user and transfers to a human agent only if it is absolutely necessary (due to requirement of actions not under agent's control) and not because of its inability to find a solution. Guidelines for task identification are as before & Whether escalation to human agent for SIM lock issue is under agent's control or not & No & Give the number of tasks successfully resolved. A task is considered resolved when the agent is able to provide a satisfactory resolution to the user and transfers to a human agent only if user's SIM is locked and needs human intervention to unlock. Guidelines for task identification are as before \\
    \hline
    The second task (unlocking SIM) is not resolved by the agent, as it requires escalation to a human agent VS The task was not fully resolved by the agent, as the user ultimately needed to be transferred to a human agent for SIM PIN/PUK unlock, which is outside the assistant's scope. Therefore, the task is not counted as successfully resolved & gpt-4.1 & Give the number of tasks successfully resolved. A task is considered resolved when the agent is able to provide a satisfactory resolution to the user and transfers to a human agent only if it is absolutely necessary (due to requirement of actions not under agent's control) and not because of its inability to find a solution. Guidelines for task identification are as before & Whether escalation to human agent for SIM lock issue is under agent's control or not & No & Give the number of tasks successfully resolved. A task is considered resolved when the agent is able to provide a satisfactory resolution to the user and transfers to a human agent only if user's SIM is locked and needs human intervention to unlock. Guidelines for task identification are as before \\
    \hline
    The agent successfully handled the task up to the point where escalation to a human agent was necessary due to SIM PIN/PUK security restrictions, which is appropriate and not a failure of the agent VS The second task (unlocking the SIM) could not be resolved by the agent due to security restrictions and was appropriately escalated to a human agent, which is acceptable as it is outside the agent's control (not given success) & gpt-4.1 & Give the number of tasks successfully resolved. A task is considered resolved when the agent is able to provide a satisfactory resolution to the user and transfers to a human agent only if it is absolutely necessary (due to requirement of actions not under agent's control) and not because of its inability to find a solution. Guidelines for task identification are as before & Whether escalation to human agent for SIM lock issue is under agent's control or not & No & Give the number of tasks successfully resolved. A task is considered resolved when the agent is able to provide a satisfactory resolution to the user and transfers to a human agent only if user's SIM is locked and needs human intervention to unlock. Guidelines for task identification are as before \\
    \hline
    The agent successfully resolved the task to the best of its ability. The issue was ultimately a SIM PIN lock, which requires human intervention for security reasons. The agent correctly identified this and transferred the user only when absolutely necessary VS The task was not fully resolved by the agent. The agent had to transfer the user to a human agent due to a SIM PIN lock, which is outside the agent's control, so the task is not counted as succeeded & gpt-4.1 & Give the number of tasks successfully resolved. A task is considered resolved when the agent is able to provide a satisfactory resolution to the user and transfers to a human agent only if it is absolutely necessary (due to requirement of actions not under agent's control) and not because of its inability to find a solution. Guidelines for task identification are as before & Whether escalation to human agent for SIM lock issue is under agent's control or not & Yes & Give the number of tasks successfully resolved. A task is considered resolved when the agent is able to provide a satisfactory resolution to the user and transfers to a human agent only if user's SIM is locked and needs human intervention to unlock. Guidelines for task identification are as before \\
    \hline
     &  &  &  &  &  \\
    \hline
    The agent followed a logical troubleshooting process, verifying account details, checking device settings, and confirming network configurations before escalating to a human agent when the issue could not be resolved through standard steps VS The agent was unable to fully resolve it and had to transfer the user to a human agent, so the task did not succeed & gpt-5-chat-latest & Give the number of tasks successfully resolved. A task is considered resolved when the agent is able to provide a satisfactory resolution to the user and transfers to a human agent only if it is absolutely necessary (due to requirement of actions not under agent's control) and not because of its inability to find a solution. Guidelines for task identification are as before & What counts as a necessary escalation to human. Cases where agent tried several steps and was unable to help were flagged as necessary human intervention on the account that advaned support from human is required & NA &  \\
    \hline
    The agent followed a structured troubleshooting process and ultimately transferred the user to a human agent after exhausting all standard steps, which is appropriate, so the taskis considered successfully resolved VS The agent followed a logical troubleshooting process, checking network, APN, permissions, and other relevant settings. However, the issue was not resolved within the agent’s control, and the user was transferred to a human agent, so the task is not considered successfully resolved & gpt-5-chat-latest & Give the number of tasks successfully resolved. A task is considered resolved when the agent is able to provide a satisfactory resolution to the user and transfers to a human agent only if it is absolutely necessary (due to requirement of actions not under agent's control) and not because of its inability to find a solution. Guidelines for task identification are as before & What counts as a necessary escalation to human. Cases where agent tried several steps and was unable to help were flagged as necessary human intervention on the account that advaned support from human is required & NA &  \\
    \hline
     &  &  &  &  &  \\
    \hline
    The agent successfully guided the user through troubleshooting steps and correctly transferred to a human agent when the SIM was locked, which is the appropriate resolution for this scenario. VS The task is not fully succeeded because the user's service was not restored within the bot's scope—the final resolution (SIM unlock) requires human intervention. Thus, the agent performed correctly, but the task remains unresolved within the automated system & gpt-4.1 & Give the number of tasks successfully resolved. A task is considered resolved when the agent is able to provide a satisfactory resolution to the user and transfers to a human agent only if user's SIM is locked and needs human intervention to unlock. Guidelines for task identification are as before &  & NA & NA \\
    \hline
    The agent successfully resolved the user's issue up to the point where human intervention is required (SIM PIN lock), which is the correct protocol VS The task was not fully resolved by the agent, as the user ultimately needed to be transferred to a human agent due to a SIM PIN lock, which the agent could not resolve & gpt-4.1 & Give the number of tasks successfully resolved. A task is considered resolved when the agent is able to provide a satisfactory resolution to the user and transfers to a human agent only if user's SIM is locked and needs human intervention to unlock. Guidelines for task identification are as before &  & NA & NA \\
    \hline
\label{tab:misclassify}
\end{longtable}

\end{document}